\documentclass[manuscript,screen]{acmart}
\AtBeginDocument{%
 \providecommand\BibTeX{{%
 \normalfont B\kern-0.5em{\scshape i\kern-0.25em b}\kern-0.8em\TeX}}}

\copyrightyear{2024}
\acmConference[arXiv]{arXiv}{2024}{}

\begin{document}

\title{Tappy: Predicting Tap Accuracy of User-Interface Elements by Reverse-Engineering Webpage Structures}

\author{Hiroki Usuba}\email{c-hiusuba@lycorp.jp}
\affiliation{\institution{LY Corporation}\city{Tokyo}\country{Japan}}

\author{Junichi Sato}\email{jsato@lycorp.jp}
\affiliation{\institution{LY Corporation}\city{Tokyo}\country{Japan}}

\author{Naomi Sasaya}\email{nasasaya@lycorp.jp}
\affiliation{\institution{LY Corporation}\city{Tokyo}\country{Japan}}

\author{Shota Yamanaka}\email{syamanak@lycorp.co.jp}
\orcid{0000-0001-9807-120X}
\affiliation{\institution{LY Corporation}\city{Tokyo}\country{Japan}}

\author{Fumiya Yamashita}\email{fyamashi@lycorp.co.jp}
\affiliation{\institution{LY Corporation}\city{Tokyo}\country{Japan}}

\renewcommand{\shortauthors}{Usuba et al.}
\renewcommand{\shorttitle}{Tappy}

\begin{abstract}
Selecting a UI element is a fundamental operation on webpages, and the ease of tapping a target object has a significant impact on usability.
It is thus important to analyze existing UIs in order to design better ones. However, tools proposed in previous studies cannot identify whether an element is tappable on modern webpages.
In this study, we developed Tappy that can identify tappable UI elements on webpages and estimate the tap-success rate based on the element size.
Our interviews of professional designers and engineers showed that Tappy helped discussions of UI design on the basis of its quantitative metric.
Furthermore, we have launched this tool to be freely available to external users, so readers can access Tappy by visiting the website (\url{https://tappy.yahoo.co.jp}).
\end{abstract}



\maketitle

\section{Introduction}
\label{sec:intro}
Smartphone-app and website developers and human-computer interaction (HCI) researchers recognize the importance of designing user interfaces (UIs) that can be correctly used.
Since one of the most fundamental operations on smartphones is selecting a UI element (button, hyperlink, picture with a link, etc.) by tapping it, the ease of tapping a target object has a significant impact on usability.
Numerous books on UI design have thus recommended making objects on screen large enough to tap easily, and many studies support this idea \cite{Clark16book,Hoober11book,Johnson14book,Neil14book}.
From the perspective of developers, it is a liability that consumers quit their services because of difficulty with the UIs.

In order to design better UIs, it is also important to analyze existing ones.
Such analyses are common practice for developers when they design their own products.
In addition, there are helpful guidelines on the target sizes on touchscreens, e.g., the Android design guideline recommends 9 mm or larger \cite{Android23}.
However, there are situations where many UI elements need to be displayed on a single screen or webpage, which would restrict the target size.
In such cases, quantitative models that estimate the success rate of tapping based on the target size would be useful \cite{Bi16,Yamanaka20issFFF}, but these models would not work well when the tappable area is different from the apparent size \cite{Usuba21iss}.

Our contributions are twofold.
First, we implemented a tool called ``Tappy'' that identifies tappable UI elements on webpages for smartphones and estimates the tap success rate.
Second, we enabled members of Yahoo Japan Corporation to use Tappy and interviewed professional designers and engineers who analyzed their products' webpages.
They found several elements with low success rates and discussed revised designs on the basis of what Tappy revealed, which is evidence that Tappy helped in their decision making.

Although HCI researchers have claimed that human-motor performance models (e.g., Fitts' law \cite{Fitts54}, success-rate models \cite{Bi16,Usuba22iss}) are useful for designing UIs \cite{Yamanaka20issFFF,Zhang23shape,Huang20cross}, we were unable to find any papers reporting how they develop and then release a tool that utilizes such models for supporting professional designers' work.
Our study is the first to do so; we have released Tappy to the industry and the professionals benefited from it.
In addition, we have launched a website tailored for external users to access Tappy, which became publicly available on January 31, 2024 (\url{https://tappy.yahoo.co.jp}).

\section{Related Work}
\label{sec:RW}
\subsection{Objectives to Analyze Webpages}
The general importance of UI design was explained in Section~\ref{sec:intro}.
More specifically, in Section~\ref{sec:RW}, we explain methods and tools to analyze UIs.
The purposes of analyzing UIs include finding the optimal position for inserting advertisements \cite{Wu13} and detecting phishing attacks by analyzing document object models (DOMs) \cite{Mao18}.
While some of these methods are inspiring, their detailed results (e.g., accuracy of phishing-attack detection) are outside the scope of our study.
For more information on how to analyze webpages systematically, readers are directed to a recent survey \cite{Prazina23}.

By analyzing DOMs, researchers can implement a tool to improve target-selection operations.
For example, there is an extension for the Chrome browser \cite{bubbleChrome} to perform the \textit{Bubble Cursor} technique \cite{Grossman05}.
However, according to its implementation, it detects the clickable area by the name and attributes of the HTML tags \cite{bubbleChromeTag}.
Thus, this extension cannot capture all elements where a click event is registered (see Section~\ref{sec:implementation} for details).

\subsection{Reverse Engineering to Obtain Source Codes}
Shirazi et al. reverse-engineered 400 Android apps to obtain the original program codes \cite{Shirazi13}.
They reported common practices such as typical screen structure consisting of three nested layers.
Silva conducted similar analyses of web applications in order to reproduce screen structures by reverse-engineering source codes \cite{Silva12}.
These studies reported examples where source codes could not be successfully generated, while Tappy does not need original codes.

\subsection{Visual Appearance-Based UI-Element Recognition}
Instead of disassembling source code, several researchers have analyzed screen structures from their appearance.
REMAUI is a tool to analyze screenshots by using OCR and computer vision techniques to identify UI elements such as images and texts \cite{Nguyen15}.
Their estimated results had an accuracy of $\sim$95\% relative to the ground truth.

Doosti et al. analyzed smartphone apps' view hierarchies and screenshots to automatically detect screen-composition elements \cite{Doosti18}.
This does not work if instances are implemented in a non-standard way.
Using convolutional neural networks (CNNs) partially resolved this issue and they achieved an accuracy of 95\%.

Schoop et al. achieved to predict whether elements in a mobile UI screenshot can be selected (referred to as \textit{tappability}) by a deep learning-based approach \cite{SchoopCHI2022}.
The precision was 91\% and better than a previous study \cite{SwearnginCHI2019}.
Wu et al. improved this method to implement a tool that estimates the tappability by automatically interacting with screens \cite{Wu23}.

\subsection{Combining Systematic and Human-Powered UI-Element Detections}
The Rico dataset on mobile apps has frequently been used \cite{Deka17Rico}.
The apps to be analyzed are provided to crowd workers, and when they use the apps, a system records their operations and identifies the UI elements.
An automated agent then continues the operations and identifies more UI elements. However, this approach incurs costs and a time lag, which are disadvantages when designers want to analyze apps or webpages.
Liu et al.'s work is an example of using the Rico dataset \cite{Liu18}.
They trained a CNN to classify UI elements and achieved an accuracy of 94\% in distinguishing icon classes.

\subsection{Prefab Toolkits}
Dixon et al. developed a series of Prefab toolkits, with which UI elements are analyzed to determine whether each pixel senses a click event \cite{Dixon10prefab,Dixon11prefab,Dixon12prefab}.
Prefab detects clickable areas by detecting pixel-based features of widgets (e.g., OK/Cancel dialogs, minimize/maximize/close buttons of windows) that have a standard visual appearance.

In contrast, we focus on webpages whose clickable/tappable UI elements are not necessarily designed to have a consistent appearance.
For example, it is sometimes impossible to determine from appearances whether a certain text has a link or whether the area around the text is also tappable.
To address this issue, Dixon et al. developed an annotation tool to identify such selectable areas, but this manual approach is ad-hoc and time-consuming \cite{Dixon14prefab}.

In summary, there are methods to automatically identify clickable/tappable areas by analyzing the visual appearance and DOM of a webpage for which the source code is not available.
However, regarding usability analysis, the detection accuracy, processing time and cost cannot be neglected.
Without source code, we do not have a tool that can automatically identify tappable areas with an accuracy of more than 95\%.
In addition, it is unclear whether the tools proposed in the previous studies are accurate on modern webpages and whether providing such facilitation tools can actually help professional designers find usability issues; this state of affairs motivated us to conduct this work.

\section{Tappy}
\begin{figure}[t]
 \centering
 \includegraphics[width=1.0\textwidth]{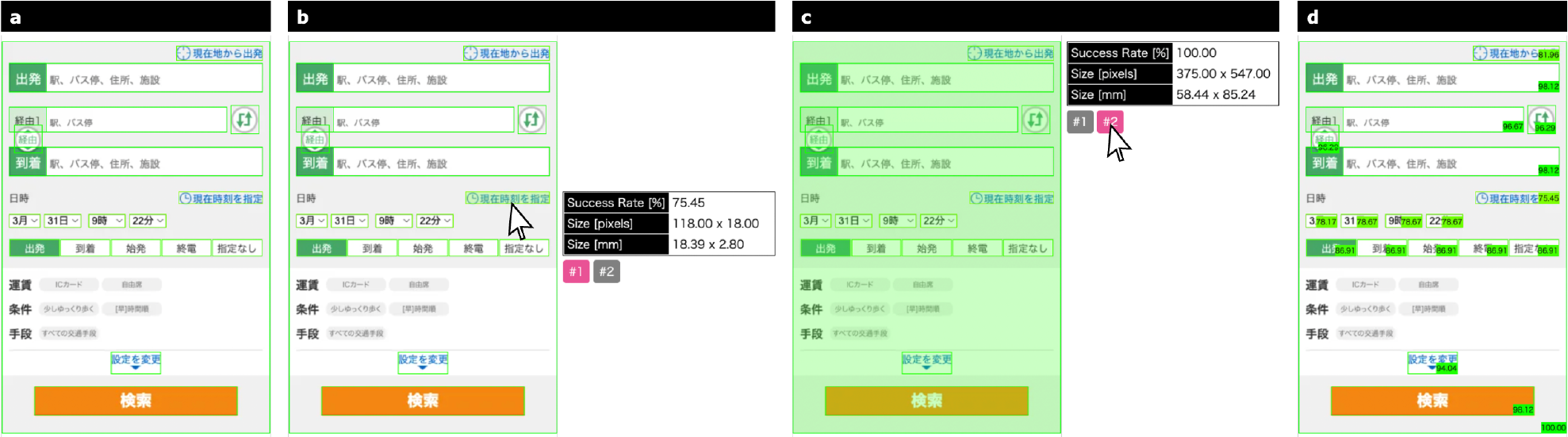}
 \caption{Prediction results of Tappy for \url{https://transit.yahoo.co.jp/}. (a) Each tappable element is surrounded by a rectangle. (b) When users select an element, the success rate, pixel size, physical size, and candidate overlapped elements are displayed. (c) Users can switch to the element that they wish to analyze by selecting the candidate. (d) The success rates can be displayed on the elements. Note that the screenshot in this figure has been clipped; the actual screenshot continues below it.}
 \label{figures/tappy}
\end{figure}

Tappy is a web application that presents the success rate of tapping each tappable element on a mobile webpage (Fig.~\ref{figures/tappy}).
Users enter the URL of a webpage and select \textit{options} (e.g., a smartphone device model; see Section~\ref{sec:tappy-options}).
Then, Tappy returns a screenshot of the webpage (Fig.~\ref{figures/tappy}a) where each tappable element is surrounded by a rectangle.
The rectangle color (red--green) corresponds to the success rate (0--100\%).
When the users select a tappable element, Tappy displays the success rate, pixel size, physical size, and candidate elements (Fig.~\ref{figures/tappy}b).
In a case where two or more elements overlap at the same cursor position, Tappy displays all candidates (items labeled ``\#1'' and ``\#2'' in Fig.~\ref{figures/tappy}b and c), and users can then switch to the desired element.

\subsection{Prediction of Tap Success Rate from Target Size}
According to the \textit{dual Gaussian distribution model}, the tap-point variance $\sigma^2$ and the square of target size are linearly related for each of the x- and y-axes independently \cite{Bi13a,Bi16}.
Assuming that the tap coordinates are random variables that obey normal distributions \cite{Bi13a,Yamanaka23HPO}, Tappy uses Usuba et al.'s model that predicts the probability that a tap falls within a rectangular target, i.e., the success rate $\mathit{SR}$, as follows \cite{Usuba22iss}:
\begin{equation}
 \label{eqs:srDefined}
 \mathit{SR} = \text{erf}\left( \frac{W}{2 \sqrt{2} \sigma_x} \right) \text{erf}\left( \frac{H}{2 \sqrt{2} \sigma_y} \right),\ \ \mathrm{where}\ \ \ \sigma_x = \sqrt{0.007101 W^2 + 1.412} \ \ \text{and}\ \ 
 \sigma_y = \sqrt{0.01181 H^2 + 1.365},
\end{equation}
where $W$ and $H$ are the target sizes on the x- and y-axes in mm, respectively.
Tappy uses the necessary coefficient values shown in Eq.~\ref{eqs:srDefined} from the data of Usuba et al.'s crowdsourced user experiment \cite{Usuba22iss} to predict $\sigma_x$ and $\sigma_y$.

\subsection{Five Options in Tappy}
\label{sec:tappy-options}

\textbf{Device name}: users can select a smartphone model from a list. Tappy simulates the size, pixels per inch (PPI), and user agent of the selected model. This selection is required. Currently, Tappy lists only iPhone models because the specs of those models have been officially compiled. However, if we had the specs of, e.g., Android smartphones and Microsoft Surface PCs, Tappy could technically  simulate such devices.

\textbf{Waiting time}: this is the duration from when Tappy completes loading a webpage to when it starts predicting the success rates. Sometimes, elements are dynamically created by JavaScript. In this case, if Tappy gets the elements immediately after the webpage completed loading, it may miss elements that have not been rendered. Thus, we set a certain waiting time. The default value is 3000 ms, which is determined by heuristics: we confirmed that Tappy correctly gets all the elements in many webpages with this setting.

\textbf{Executing JavaScript}: if this option is enabled, Tappy loads the scripts and executes them. As mentioned above, the elements are sometimes dynamically created by JavaScript. For such elements, Tappy needs to execute JavaScript. However, users sometimes may wish to analyze a webpage before JavaScript is executed, e.g., when advertisements created in JavaScript appear over elements. In this case, users disable this option. That is, the default setting is enabled.

\begin{figure}[t]
 \centering
 \includegraphics[width=0.47\textwidth]{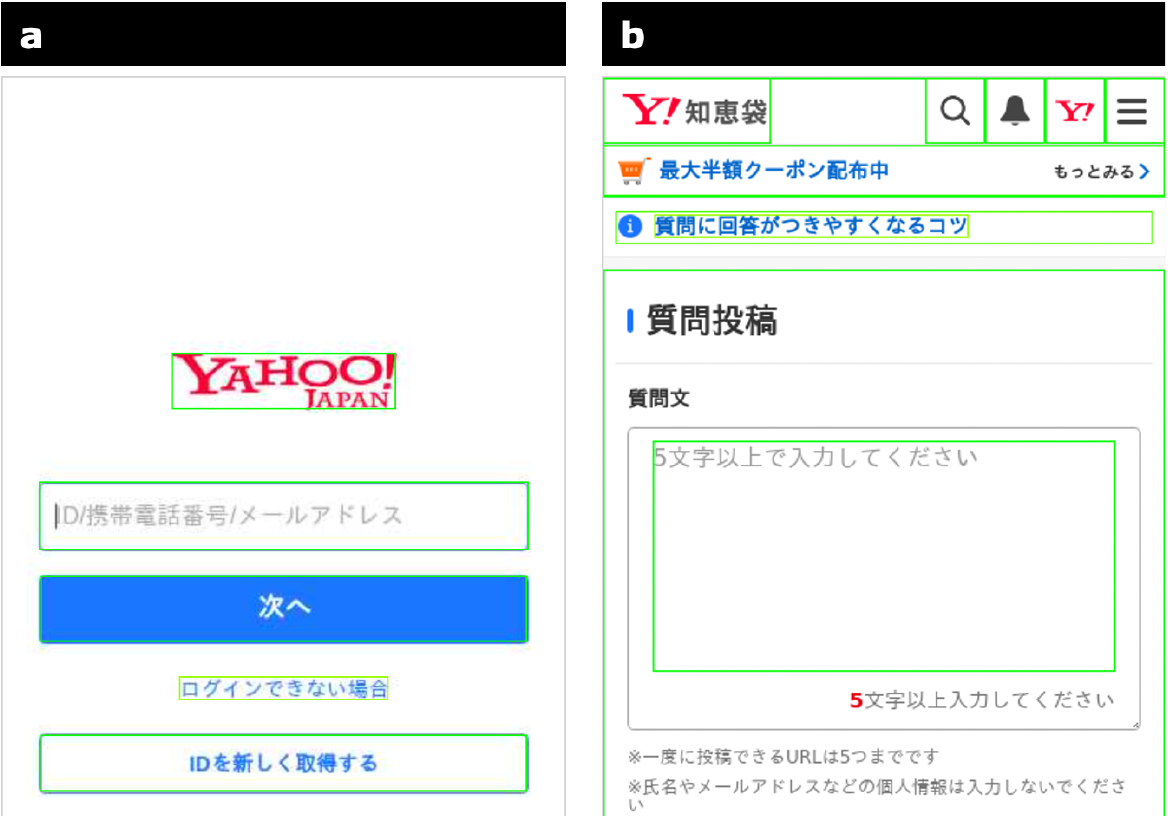}
 \caption{Prediction results of Tappy for \url{https://chiebukuro.yahoo.co.jp/question/post/form}. (a) When users are not logged in, the page shows the login form. (b) When users log in, the page shows the question submission form.}
 \label{figures/cookie}
\end{figure}

\textbf{Sending cookies}: if this option is enabled, users can pass web cookies to Tappy. Here, for instance, users cannot access a ``Cart'' page on a shopping site if they do not log in. The login information is typically saved in the cookies after the users log in, and Tappy can access such pages by receiving the cookies. The users can pass the cookies to Tappy by confirming the cookies with (e.g.,) Chrome DevTools and pasting the values in a form in Tappy. The cookies may include personal information. Thus, the default is disabled, and if the users wish to enable this option, they must approve that Tappy logs in on behalf of them. If this option is enabled, we do not save a screenshot, HTML, and the cookies.
Figure~\ref{figures/cookie} shows that the pages (a) before and (b) after logging in. On this website, the users can post questions, like in Quora \cite{Quora}, but they must be logged in.

\textbf{List success rates}: if this option is enabled, Tappy displays all the success rates over the elements (Fig.~\ref{figures/tappy}d).

\subsection{Implementation}
\label{sec:implementation}
Tappy receives the URL that the users enter and the above options, and it accesses the URL with Headless Chrome (Puppeteer \cite{Puppeteer}). Thus, Tappy does not analyze the webpage (e.g.,) by parsing the HTML DOMs, but does render the webpage. After Tappy completes loading the URL and executing JavaScript, it gets all \textit{tappable} elements. The tappable elements are of two types: (1) elements having specific names and (2) elements having event listeners. In addition, tappable elements exist (3) in \texttt{<iframe>}.

\textbf{(1) Elements having specific names}: in HTML, for elements rendered as tappable objects, there are \texttt{<a>}, \texttt{<button>}, \texttt{<input>}, \texttt{<select>}, \texttt{<textarea>}, and \texttt{<label>} (for \texttt{<input>}). Such elements can be gotten with \texttt{querySelectorAll} \cite{QuerySelectorAll} whose query consists of the tag names.

\textbf{(2) Elements having event listeners}: the users can tap elements where the event listeners are registered. If the event listeners are registered by the event attributes (e.g., \texttt{onclick}), the elements can be gotten by the selector whose query consists of the attribute names. However, if the event listeners are registered by \texttt{addEventListener} \cite{AddEventListener}, the elements cannot be gotten by any vanilla JavaScript API, because there are no APIs confirming the event listeners that elements have. However, in the Chrome console, \texttt{getEventListeners} \cite{GetEventListeners}, which lists the event listeners that the elements have, can be used, and this method can be used in Puppeteer. Thereby, Tappy gets elements that have Touch Events \cite{TouchEvents}, Pointer Events \cite{PointerEvents}, and/or UI Events \cite{UIEvents} (excluding the abort, error, load and unload events not fired from the users' actions) listeners.

\textbf{(3) Elements in \texttt{<iframe>}}: advertisements often are embedded as \texttt{<iframe>} in HTML. Basically, because the advertisements have different domains from that of the embedded webpage, the elements in \texttt{<iframe>} cannot be gotten with vanilla JavaScript APIs due to the same-origin policy, which prevents JavaScript from accessing sources that have different domains. However, Tappy can get the elements in \texttt{<iframe>} with Chrome DevTools Protocol \cite{ChromeDevToolsProtocol} and can get the (1) and (2) elements described above in \texttt{<iframe>}.

With the above procedure, Tappy gets all touchable elements; this cannot be achieved by simply parsing the HTML DOMs.
Because Tappy can consider some style properties (e.g., \texttt{opacity}, \texttt{visibility}), the above touchable elements do not include invisible elements.
Then, Tappy gets the pixel size and position of the elements with \texttt{getBoundingClientRect} \cite{GetBoundingClientRect}, calculates the physical sizes and positions in the specified device model, and predicts the success rates with Eq.~\ref{eqs:srDefined}.
Finally, Tappy returns a screenshot of the webpage and the above information about the elements.
The range of the screenshot is from the page top to the page bottom.
Thus, Tappy takes the screenshot while scrolling the page. However, the screenshot does not include the range loaded during scrolling. We know that some elements only appear after scrolling. Thus, technically, we can modify Tappy to scroll until all elements are rendered or to scroll for a certain amount of time. The users can see the prediction results as shown in Fig.~\ref{figures/tappy}.

\section{Interviews}
We made Tappy accessible only to Yahoo Japan Corporation's employees.
The employees used it 756 times in the first month after release.
There were 293 unique users, including professional designers and engineers, and 335 unique URLs.
We sent a questionnaire to Tappy's users (e.g., \textit{How was the usability of Tappy?}, \textit{In what situations did you use Tappy?}).
Then, we interviewed the users to ask them about Tappy usage examples.
Our affiliation's IRB-equivalent ethics team approved this study.

\subsection{User~1: Analyzing Yahoo! JAPAN Travel}
User~1 analyzed Yahoo! JAPAN Travel \cite{YJTravel}, selecting an iPhone~12 mini as the device model. The user commented that Tappy showed that the success rate for the hamburger icon was too low (58\%, Fig.~\ref{figures/travel}), whereas the rates of almost all other tappable objects were $\geq$ 98\%.
If designers try improving the usability, because there are margins around the hamburger icon, the success rate can be increased by adding \texttt{padding} \cite{Padding} without changing the icon's appearance. Considering the size of the header, the tappable width and height of the icon can be enlarged to 7.04~mm, enabling the success rate to reach $\geq$ 98.0\%.

\subsection{User~2: Analyzing Yahoo! JAPAN Search}
\begin{figure}[t]
  \begin{minipage}[b]{0.48\textwidth}
    \centering
    \includegraphics[width=1.0\textwidth]{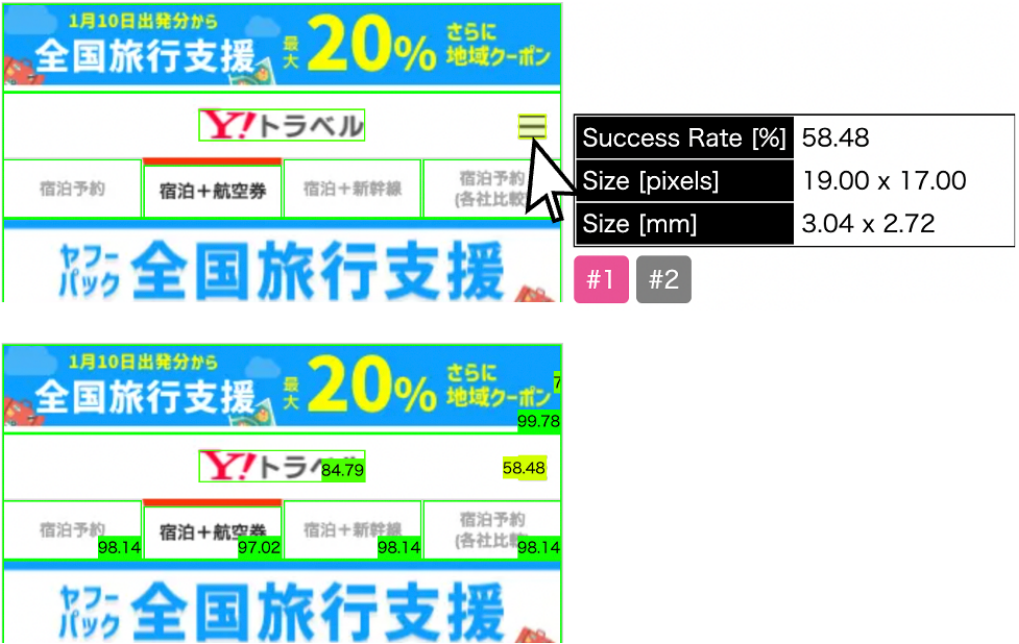}
    \caption{Prediction results for Yahoo! JAPAN Travel. The success rate for the hamburger icon is too low (58\%).}
    \label{figures/travel}
  \end{minipage}
  \hspace{0.02\textwidth}
  \begin{minipage}[b]{0.48\textwidth}
    \centering
    \includegraphics[width=1.0\textwidth]{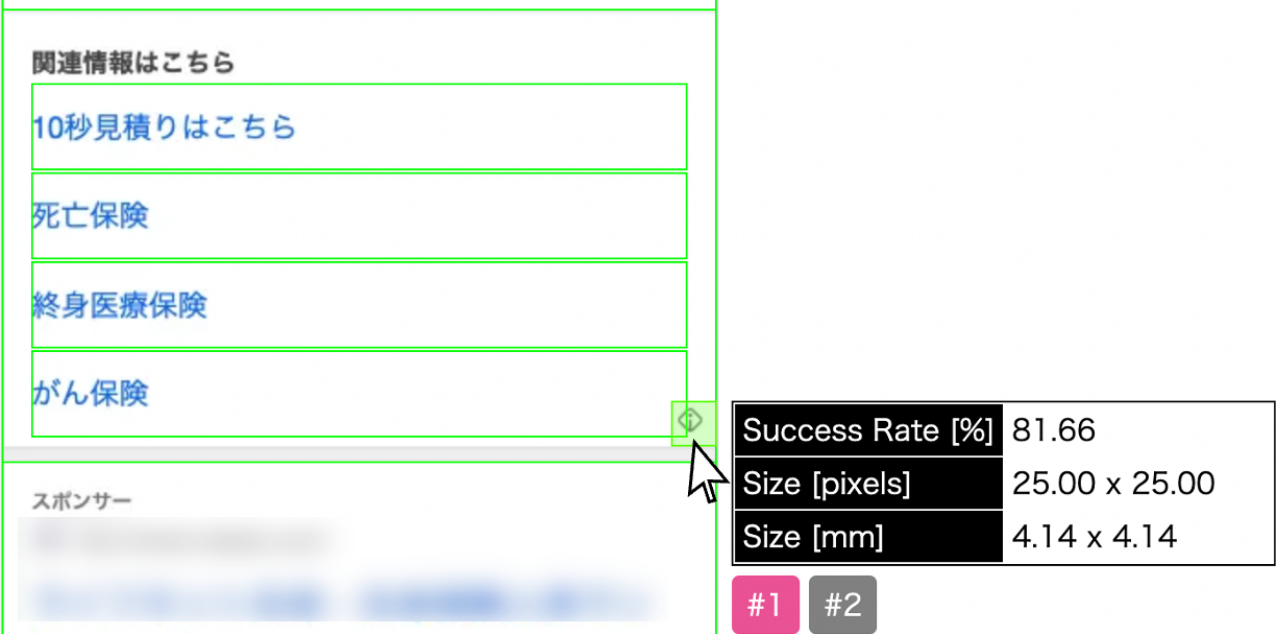}
    \caption{Prediction results for Yahoo! JAPAN Search. The success rate for the information icon is low.}
    \label{figures/information}
  \end{minipage}
\end{figure}

User~2 analyzed search results for the word ``Insurance'' in Japanese in Yahoo! JAPAN Search \cite{YJSearch} with an iPhone~13. Yahoo! JAPAN Search shows advertisements related to the search query.
Figure~\ref{figures/information} shows four links related to the advertisement with an information icon. When the information icon is selected, a popup appears to describe why the advertisement is shown. The user found that the success rate for the information icon is low (81.66\%).
Because this icon overlaps the fourth link to the related page, if the icon size remains small, an accidental tap to the link may happen frequently.
If the icon is enlarged by 3~mm, the success rate can be increased to $\geq$ 95\%. Alternatively, the number of accidental taps can be decreased by adding margins between the related-page links and the information icon \cite{Yamanaka18mobilehci}.

\subsection{User~3: Analyzing Yahoo! JAPAN Search}
\begin{figure}[t]
  \begin{minipage}[b]{0.48\textwidth}
    \centering
    \includegraphics[width=1.0\textwidth]{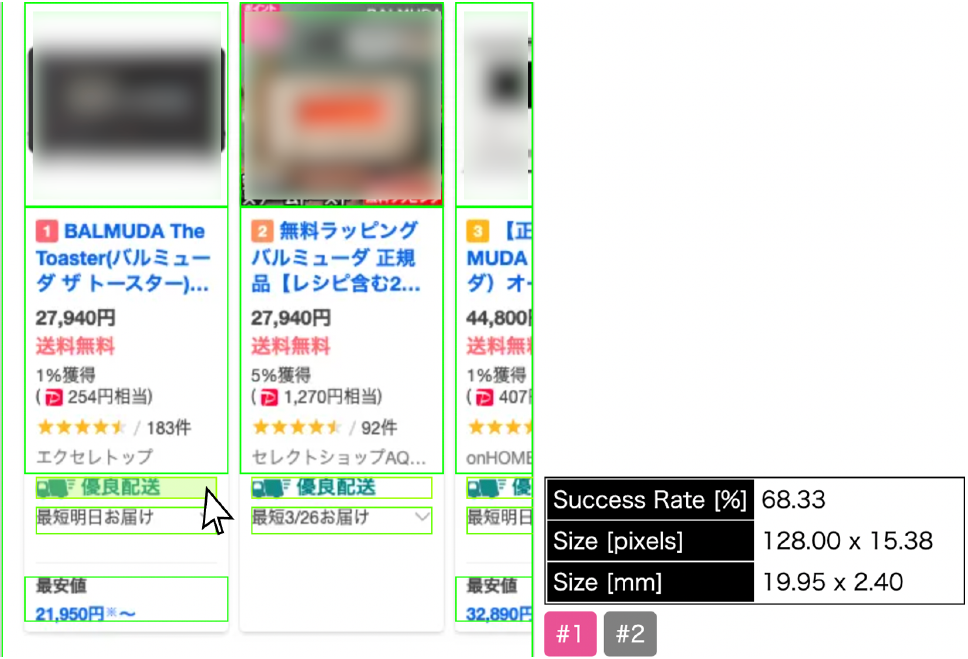}
    \caption{Prediction results for electrical applications in Yahoo! JAPAN Search. The success rate for the button is low.}
    \label{figures/balmuda}
  \end{minipage}
  \hspace{0.02\textwidth}
  \begin{minipage}[b]{0.48\textwidth}
    \centering
    \includegraphics[width=1.0\textwidth]{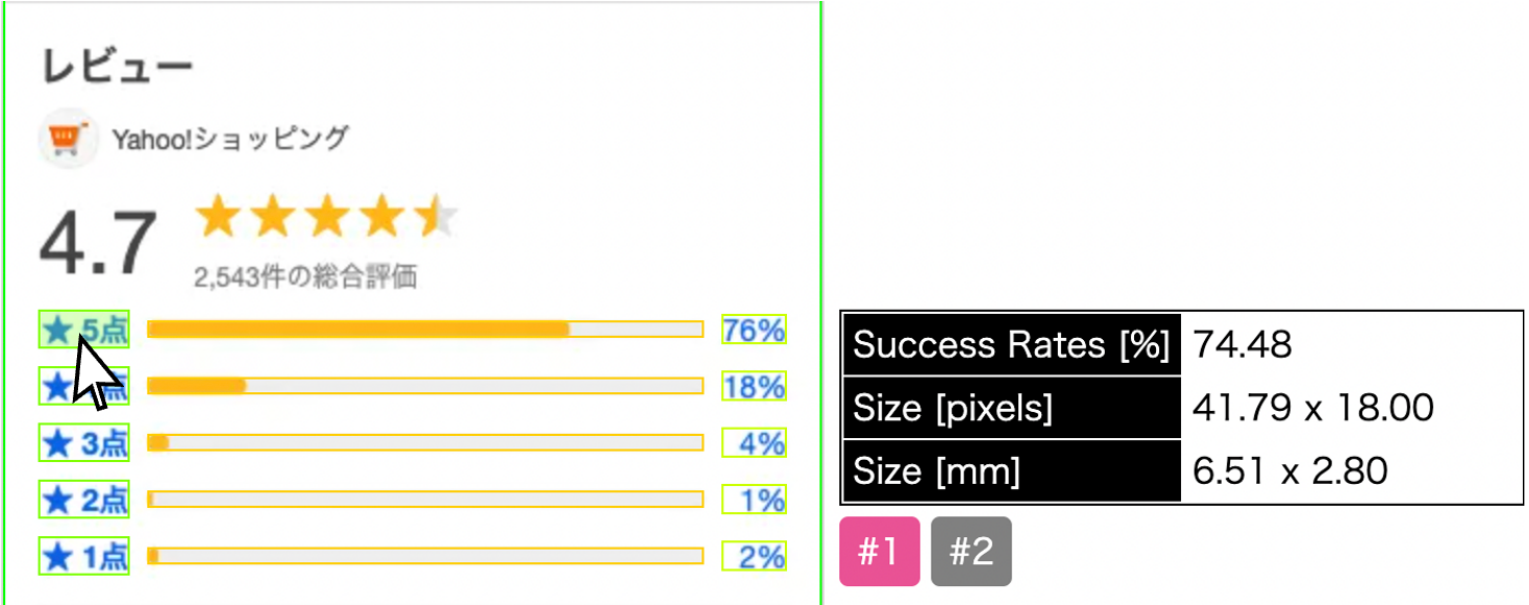}
    \caption{Prediction results for ratings in Yahoo! JAPAN Search. The success rate for the button is low.}
    \label{figures/rating}
  \end{minipage}
\end{figure}

User~3 analyzed search results on the electrical appliances of a certain company in Yahoo! JAPAN Search with an iPhone~SE (3rd gen). In addition to showing general web-search results, Yahoo! JAPAN Search recommends several products sold in Yahoo! JAPAN Shopping based on the search query.
Figure~\ref{figures/balmuda} shows a page on three toaster ovens displaying the image, name, price, etc., of each.
Beneath each product, there are two buttons that show notes from the store, which were implemented by User~3.
User~3 was surprised at the low success rates of the buttons.
In addition, Yahoo! JAPAN Search also shows reviews of the electrical appliances (Fig.~\ref{figures/rating}), and User~3 found that the success rates for the rating buttons were low (74\%).
User~3 suggested to the product team that they should increase the button size.
However, the suggestion was pending because the buttons are not frequently selected and of low importance.

\begin{figure}[t]
 \centering
 \includegraphics[width=0.9\textwidth]{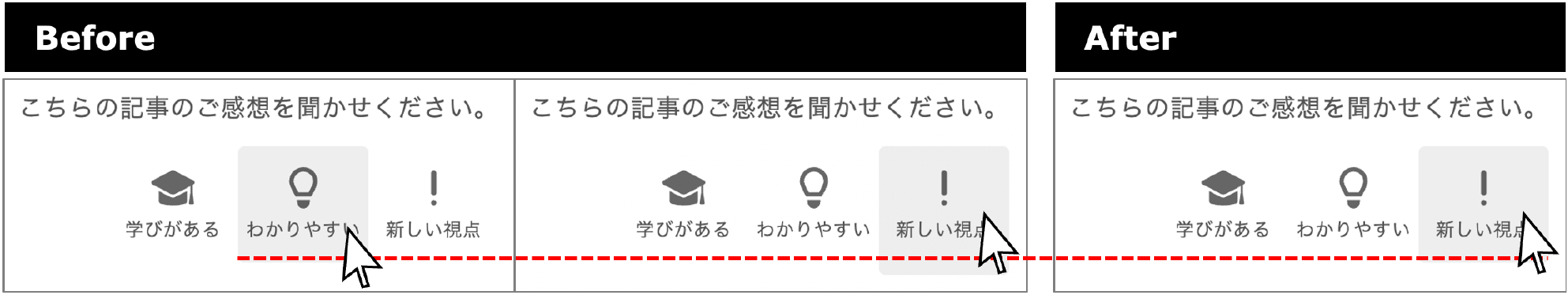}
 \caption{Before and after changing the voting buttons in Yahoo! JAPAN Tech Blog.}
 \label{figures/techblog}
\end{figure}

\subsection{User~4: Analyzing Shopping Sites}
User~4, who is a designer of a shopping site, used Tappy to compare their ``Add to cart'' buttons with those on other shopping sites.
User~4 found the button height is somewhat small in the site that User~4 is in charge of.
Thus, User~4 suggested to the product team that they should increase the button height.
However, the suggestion was rejected because there was no previous evidence that increasing the success rates had increased some sales contributions.
While, the product team evaluated that Tappy was appropriate as a verification tool.
User~4 commented that Tappy becomes better if it is clear that which sales metrics are improved by increasing the success rates.

\subsection{User~5: Analyzing Yahoo! JAPAN Tech Blog}
Yahoo! JAPAN Tech Blog \cite{YJTechBlog} has articles describing the technology, design, and culture supporting Yahoo! JAPAN, and the readers can cast votes by selecting from ``useful'', ``easy to understand'', or ``fresh perspective'' (Fig.~\ref{figures/techblog}). Selectable areas of the voting buttons are highlighted in gray only while tapping them, i.e., the selectable areas are basically invisible.
By analyzing the blog with Tappy, User~5, who is a designer, found that the height of the last item (``fresh perspective'') in the voting buttons is larger than the others because Tappy surrounds the selectable areas. User~5 commented that User~5 noticed a cascading style sheet (CSS) misconfiguration for the first time because Tappy visualized the selectable areas. This example is not related to the success rate, but this is surely an example illustrating the usefulness of Tappy.

\subsection{Summary of Interviews}
Some users commented that Tappy helped them to discuss webpage designs with members of the product team on the basis of the quantitative values it provided. This has never been achieved with any tools before.
Thus, we conclude that Tappy is able to decrease non-objective decision making based on (e.g.,) a few designers' heuristics.

In summary, Tappy gave users the opportunity to review their products' webpages and made them aware of low success-rate elements, which prompted them to modify these webpages.
Moreover, we found that such low success-rate buttons were designed to be small in size because they are not important or not frequently selected. The designers typically determine the size of the button by looking at the overall balance and considering the screen size.
Tappy contributes by providing the quantitative metric of tap success rate when they determine that balance.

\section{Limitations and Future Work}
The prediction model of Tappy uses data collected from a crowdsourced experiment where participants held the device with the non-dominant hand and tapped with the index finger of the dominant hand \cite{Usuba22iss}.
Thus, the actual success rate may differ from the predicted value if users hold the device with the dominant hand and tap with the thumb of that hand, but this finger-related difference in success rates should be slight \cite{Bi13b}.
We can overcome this issue by collecting additional data for other potential operation styles.
Previous studies have also shown that the success rate depends on the user’s age \cite{Findlater13age,Findlater20age,Hourcade08age}.
We thus plan to add options for ``age'' and ``how to hold the device'' after collecting the data.
In addition, although the success rate changes when the screen edges \cite{UsubaISS2023} and the distractors \cite{Yamanaka18mobilehci,YamanakaISS2019} are close to the target, Tappy's success-rate model cannot consider these factors.
Thus, we will improve the model and the success-rate prediction.

Tappy considers the shapes of the elements always to be rectangles because we wanted to make the implementation simple and obtained the shapes by using \texttt{getBoundingClientRect}. In HTML, the shapes of elements can be arbitrary, such as a diamond or an ellipse. Tappy can be made to handle the shapes of the elements more precisely by considering CSS properties and shape paths for scalable vector graphics (SVG).
Models on success rate for circular targets \cite{Yamanaka20issFFF, Bi16} and arbitrary shapes \cite{Zhang20moving} would benefit future improvements to Tappy.

Tappy uses Headless Chrome running on a Linux server for simulating a webpage. Thus, there may be differences from an actual webpage on a smartphone, but the differences would be negligibly small.
Modern webpages return a webpage for smartphones by identifying a user agent, or they automatically change it to fit the screen size (i.e., responsive page).
Thus, in devices that have the same user agent and screen size, the rendering results are almost the same.
In devices that have different HTML rendering and JavaScript engines, the rendering results are different, but the differences may be almost incomprehensible even when checking visually.
Tappy sends a user agent and device information (e.g., screen size) to Headless Chrome.
Thus, the rendering results are almost always correct.

\section{Conclusion}
We developed Tappy, which can identify tappable UI elements on webpages and estimate the tap-success rate based on the element size.
Because it is difficult to identify if an element is tappable on modern webpages, implementing a tool like Tappy has been challenging.
Our interviews demonstrated that Tappy was beneficial for professional designers and engineers.
For example, it enabled them to identify elements with low success rates and to notice implementation errors like CSS misconfigurations by visualizing the selectable areas of these elements.

\bibliographystyle{ACM-Reference-Format}
\bibliography{sample-base}

\end{document}